\documentclass[manuscript]{aastex}

\usepackage{graphicx}
\usepackage{rotating}
\usepackage{pdflscape}
\usepackage{color}
\usepackage{lineno}
\linenumbers*[1]
\graphicspath{{NewFIGS2/}}
\usepackage{mathtools}
\usepackage{amsmath}
\usepackage{float}
\usepackage{color}
\newcommand{\blue}{\textcolor{black}}
\newcommand{\red}{\textcolor{black}}
\newcommand{\beq}{\begin{equation}}
\newcommand{\eeq}{\end{equation}}
\newcommand{\baq}{\begin{eqnarray}}
\newcommand{\eaq}{\end{eqnarray}}

\shorttitle{Observation of a CIR pair event}
\shortauthors{Z. Wu, Y. Chen, G. Li, et al.}

\begin{document}
\title{Observation of Energetic particles between a pair of Corotating Interaction Regions}
\author{Z. Wu$^{1}$, Y. Chen$^{1}$, G. Li$^{2,*}$, L. L. Zhao$^{2}$,
R. W. Ebert$^{3}$, M. I. Desai$^{3,4}$, G. M. Mason$^{5}$,
B. Lavraud$^6$, L. Zhao$^{7}$,  Y. C.-M. Liu$^{8}$,
F. Guo$^{9}$, C. L. Tang$^{1}$, E. Landi$^{7}$, J. Sauvaud$^6$ }

\affil{  $^1$ Institute of Space Sciences and School of Space Science and Physics,
Shandong University at Weihai, Weihai, China 264209 \\
$^2$ Department of Space Science and CSPAR, University of Alabama in Huntsville 35899, USA  \\
$^3$ Southwest Research Institute, San Antonio, TX, USA \\
$^4$ University of Texas at San Antonio, San Antonio, TX, USA \\
$^5$ Applied Physics Laboratory, Johns Hopkins University, Laurel, MD, USA \\
$^6$ Institut de Recherche en Astrophysique et Plan\'{e}tologie, Universit\'{e} de Toulouse (UPS), Toulouse, France.
and  Centre National de la Recherche Scientifique, UMR 5277, Toulouse, France \\
$^7$ Department of Atmospheric, Oceanic, and Space Sciences, University of Michigan, Ann Arbor, MI, USA, 48105 \\
$^8$ State Key Laboratory of Space Weather, National Space Science Center, CAS. Beijing, China 100190 \\
$^9$ Theoretical Division, Los Alamos National Laboratory, NM, USA \\
$^*$ gang.li@uah.edu }

\begin{abstract}
\red{We report observations of the acceleration and trapping of energetic ions and electrons
between a pair of corotating interaction regions (CIRs). The event occurred in Carrington Rotation 2060.
Observed at spacecraft STEREO-B,  the two CIRs were separated by less than $5$ days.
In contrast to other CIR events, the fluxes of energetic ions and electrons in this event reached their maxima between the
trailing-edge of the first CIR and the leading edge of the second CIR.
The radial magnetic field ($B_r$) reversed its sense and the anisotropy of the flux also changed from sunward to
anti-sunward between the two CIRs. Furthermore, there was an extended period of counter-streaming suprathermal electrons
between the two CIRs.} Similar observations for this event were also obtained for ACE and STEREO-A.
We conjecture that these observations were due to a ``U-shape'' large scale magnetic field topology
connecting the reverse shock of the first CIR and the forward shock of the second CIR.
Such a disconnected U-shaped magnetic field topology may have formed due to magnetic reconnection in the upper corona.
\end{abstract}

\maketitle

%62767193

\section{Introduction}
\par

Corotating interaction regions (CIRs) are commonly found during solar minimum. They are formed when fast solar wind
(often from coronal holes) overtakes slow solar wind (often from near a streamer belt)
\citep{Krieger1973, Feldman1981, Gosling1981}.
Because the slow and fast winds are two plasmas of different magnetic origin, there is a stream
interface between them and two shock waves can form with a forward shock propagating into the slow
wind and a reverse shock propagating into the fast wind (but both carried outwards by the solar wind)
\citep{Smith1976, Pizzo1985}.
The shock waves are often formed beyond 1 AU  \citep{Smith1976, Pizzo1985}.
However, recent studies showed that CIR shocks can form as close as $0.7$ AU \citep{Jian2006}.

CIRs and the associated shocks are a major source of energetic particle acceleration in the interplanetary
medium; (e.g., \citet{vanHollebeke1981, Tsurutani1982}). Indeed, during solar minimum when the coronal mass
ejection (CME) rate is low, CIRs are the dominant source of energetic particles in the interplanetary medium
at 1 AU (for recent reviews; see, e.g., \citet{Fisk1999, Gosling1999, Mason1999, Scholer1999, Richardson2004}).
These energetic particles have approximately a solar wind composition
up to a few MeV nucleon$^{-1}$, and sometimes to $10$-$20$ MeV nucleon$^{-1}$,
indicating that they are accelerated from the solar wind (e.g. \cite{Mason1999}).
However, examination of particles such as He$^+$ and $^3$He showed that they are more
abundant in the CIRs than in the uncompressed solar wind, suggesting that the interplanetary suprathermal
ion pool also contributes (e.g., \citet{Gloeckler1994, Mason2000, Mobius2002, Mason2008a}).
Furthermore, recent observations by \citet{Mason2012} found that the heavy ion abundances show a clear
correlation with sunspot number, where heavy ions are more enhanced during active periods.
Although the acceleration likely occurs at CIR shocks, how the population of seed particles is injected
into the shock remains not well understood.
Electrons are also accelerated to several hundred keV in CIRs, particularly at distances beyond 1 AU.
 \citep{Simnett1995, Simnett2003}.

The peak fluxes in CIRs often occur at the reverse and forward shocks, with fluxes at the reverse shock
generally being larger \citep{Heber1999, Mason1999}. For example, in a recent survey of 50 CIRs between 2007
February and 2008 September, \citet{Bucik2009a} found that in 90\% of the events the peak flux of energetic
helium occurred within the compressed region or the trailing edge. In the remaining 10\% of the events,
the peak flux occurred within $24$ hours of the trailing edge. \citet{Ebert2012} also confirmed this result.
Note that even if there are no shocks associated with the CIRs, as long as there is significant local
compression near the CIRs, particles are subject to acceleration \citep{Desai1998, Giacalone2002}.
Indeed, \citet{Desai1998}
showed that, for CIRs near 5 AU, the local compression ratio was a key factor in determining the
intensity of ∼1 MeV protons, regardless of whether or not the CIR was bounded by shocks.
 \citet{Ebert2012} showed a similar result for $<0.8$ Mev/nucleon Helium ions for CIRs at 1 AU.

\red{
The solar wind suprathermal electron population (in the 70 to 2000 eV range) is typically made of the ``halo''
population (generally tenuous and isotropic) and the ``strahl'', an intense beam of suprathermal electrons
aligned to the magnetic field and directed outward from the Sun. Owing to their large mean-free path,
suprathermal electrons travel freely along magnetic field lines, and are therefore used as tracers of the
large-scale heliospheric topology (e.g., \citet{Gosling1990, Kahler1994}).
Although this population is usually dominated by the outward-flowing strahl, sunward-directed counter-streaming
suprathermal electrons are also often recorded in the solar wind. These are observed on closed field lines with
both ends attached to the Sun, such as within interplanetary coronal mass ejections (ICMEs)
(e.g., \citet{Philipp1987, Gosling1987}) or newly closed field regions due to magnetic reconnection at
the heliospheric current sheet (HCS) (\citet{Gosling2006, Lavraud2009}). Counter-streaming suprathermal
electrons may also occur due to the presence of magnetic field enhancements farther from the spacecraft along
magnetic field lines, e.g., within CIRs, near their bounding shocks and to some distance downstream.
The counterstreaming population in such cases is thought to result from wave-particle interactions and
shock heating, combined with adiabatic mirroring and particle leakage into the upstream regions of the CIR
(\citet{Gosling1993, Steinberg2005, Skoug2006, Lavraud2010}).
}

Because of the wavy nature of the interplanetary current sheet, multiple CIRs can form in a single solar
rotation. Will the presence of multiple CIRs affect the particle acceleration process?
In this paper, we report, to our knowledge, the first observation of an energetic particle event associated
with a CIR pair.

\section{Observations}
In this work, we use the data from STEREO's IMPACT instrument suite \citep{Luhmann2008}.
These include protons between $101.3$ keV nucleon$^{-1}$ to $1.985$ MeV nucleon$^{-1}$
and electrons from $65$ keV to $375$ keV from the Solar Electron and Proton Telescope (SEPT)
\citep{Muller2008}; $4$-$6$ MeV protons from 16 sectors of the Low Energy Telescope (LET) \citep{Mewaldt2008};
$20$ keV nucleon$^{-1}$ to $3.6$ MeV nucleon$^{-1}$ proton, Helium, Oxygen, and Fe ions from the
Suprathermal Ion Telescope (SIT) \citep{Mason2008b}; and $\sim 250$ eV electrons from the
Solar Wind Electron Analyszer (SWEA) \citep{Sauvaud2008}. We also used
the Plasma and Suprathermal Ion Composition (PLASTIC) sensor \citep{Galvin2008} for solar wind proton parameters
including the speed, number density, and total pressure. The solar wind
magnetic field is from the IMPACT/MAG instrument \citep{Acuna2008}.
We also used data from ACE/MAG \citep{Smith1998}, ACE/SWEPAM \citep{McComas1998}, and ACE/EPAM \citep{Gold1998}.

We define a CIR pair as two CIRs such that the end of the first CIR and the beginning of the second
CIR are within $5$ days. A total of $24$ CIR pairs were identified in the two-year period of 2007 and 2008.
The CIR pair we examine in this work occurred during Carrington Rotation (CR) 2060.

\begin{figure}[ht]
\includegraphics[width=0.95\textwidth]{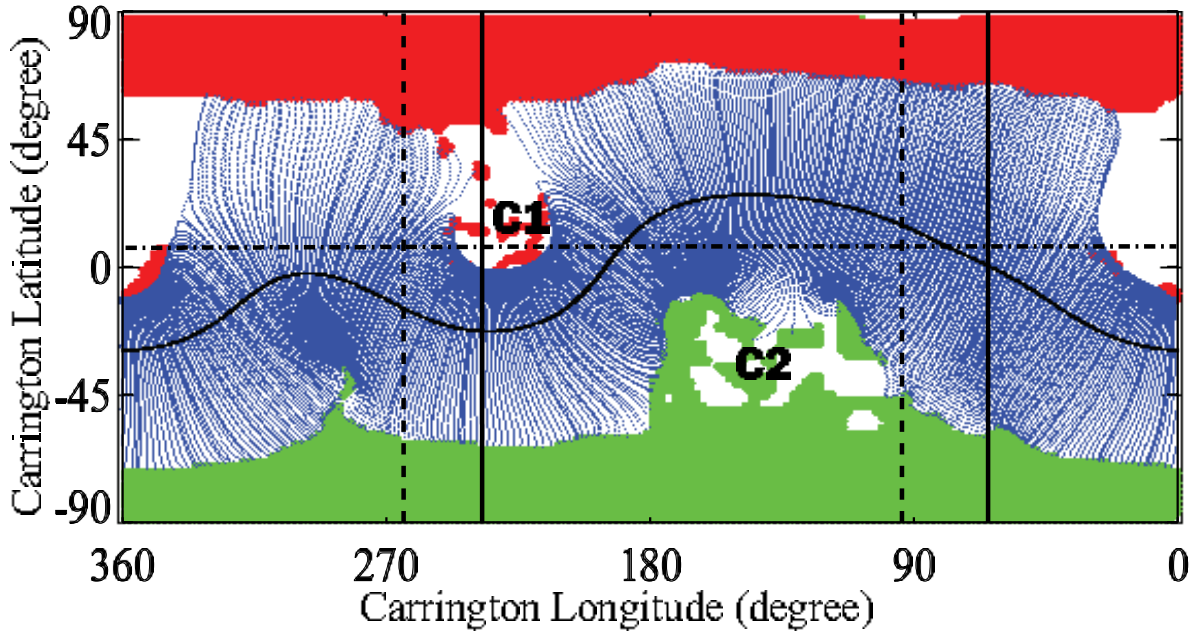}%{Fig2_New.png}
\caption{Trajectories of STEREO-A, STEREO-B and ACE, which
is the horizontal line overplotted on the Gong's synoptic map for the
Carrington Rotation 2060.   }
\label{fig:trajectory}
\end{figure}

Figure~\ref{fig:trajectory} shows the trajectories (the dot-dashed
curve above the equator) of the three spacecraft (STEREO-B, ACE
and STEREO-A) overplotted on the GONG PFSS magnetic extrapolation,
with open field footprints in red/green, closed field at the
source surface in blue, and heliospheric current sheet (HCS) at
the source surface in black (courtesy of NSO/GONG) for the
Carrington rotation 2060. The longitudinal separation between
STEREO-A and STEREO-B was $\sim 28^{\circ}$. The solid vertical
lines mark the beginning and ending longitudes for STEREO-B. The
dashed vertical lines mark the beginning and ending longitudes for
STEREO-A. ACE was located half way between STEREO-A and STEREO-B.
The coronal sources for the fast streams of the two CIRs are
labeled as ``C1'' and ``C2'' in this plot.

\begin{figure}[ht]
\includegraphics[width=0.8\textwidth]{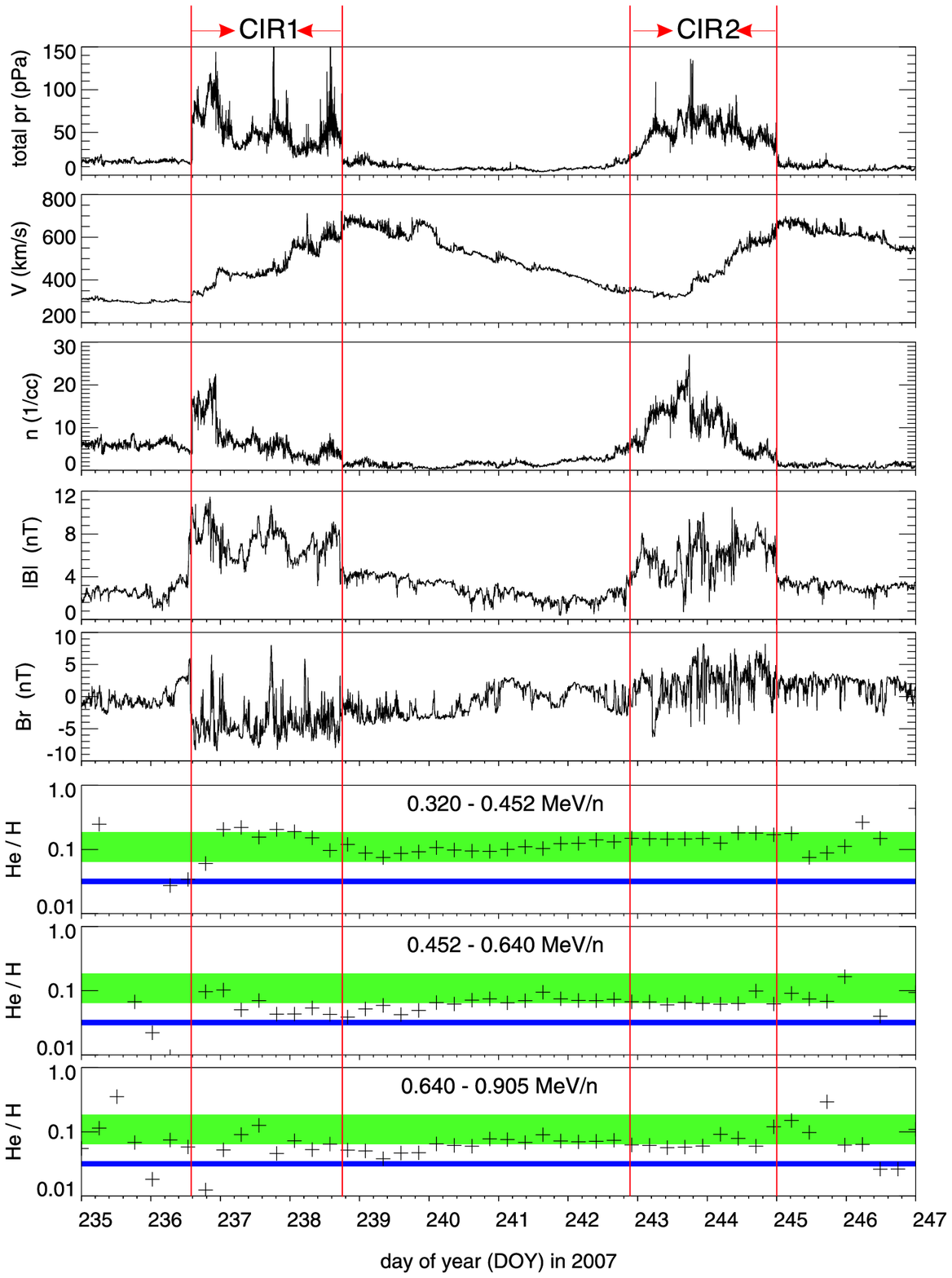}%{Fig1.png}
\caption{STEREO-B observations showing from top to bottom:
plasma total pressure, solar wind speed, total magnetic field strength,
radial magnetic field component, plasma number density,
%the pitch angle distribution (PAD) of suprathermal electrons (electrons),
and the ratio of helium to proton at
$0.320$-$0.452$ MeV nucleon$^{-1}$, $0.452$-$0.640$ MeV nucleon$^{-1}$ and $0.640$-$0.905$ MeV nucleon$^{-1}$
are shown. }
\label{fig:CIR_Observation}
\end{figure}

STEREO-B {\it in situ} observations of the two CIRs from  spacecraft
are shown in Figure~\ref{fig:CIR_Observation}.
\blue{From top to  bottom, the panels in Figure~\ref{fig:CIR_Observation} are:
 the proton pressure, the solar wind speed, the plasma number density, the solar wind magnetic field strength,
the radial component of the magnetic field,
%the pitch angle distribution (PAD) of suprathermal electrons,
and the ratio of helium to proton fluxes at $0.320$-$0.452$ MeV nucleon$^{-1}$,
$0.452$-$0.640$ MeV nucleon$^{-1}$ and $0.640$-$0.905$ MeV nucleon$^{-1}$.}

As shown in Figure~\ref{fig:CIR_Observation},
the first CIR (hereafter CIR1) occurred from 2007 August 24 (DOY 236), 14:16UT to
2007 August 26 (DOY 238), 18:10UT. The second CIR (hereafter CIR2) occurred from 2007
August 30 (DOY 242), 20:40UT to 2007 September 02 (DOY 245), 00:10UT.
We note that this event has been reported as  event No. 14 in Table 1 of \citet{Mason2009}.
Since there was only one peak in the ion time intensity profile, \citet{Mason2009}
associated this with a single CIR. However,
an examination of the solar wind profile shows that there were two CIRs.
\citet{Lavraud2010} also listed them as two separate CIRs.
Note that the separation angle between STEREO-A and STEREO-B at CIR mid-point at ACE for
this event was $28.3^{\circ}$ \citep{Mason2009}, so both CIRs were observed by all three
spacecraft.

We use sudden changes in total pressure and magnetic field to identify the boundaries (so they may
correspond to the shocks) of the two CIRs. These are shown as the red vertical lines in
Figure~\ref{fig:CIR_Observation}. CIR1 had both a forward shock and a reverse shock as can
be seen from the STEREO-B observations \footnote{note that the online catalog maintained by L. Jian
\url{http://www-ssc.igpp.ucla.edu/forms/stereo/stereo\_level\_3.html}  did not list the reverse shock of
CIR1 as a shock}.
For CIR2, the reverse shock had completely formed at STEREO-B, but the forward shock had not.

\blue{ The green and blue horizontal bars in the bottom three panels represent the typical ranges of
helium to proton ratio found in CIRs (0.125$\pm$0.061) and SEPs (0.032$\pm$0.003),
respectively \citep{Mazur1993, Mason1997, Bucik2009b}.  These He/H ratios as well as the heavy ion
ratios (not shown) clearly identify these energetic particles as being from a CIR. }

Using the CDAW Data Center catalog (\url{http://cdaw.gsfc.nasa.gov/CME\_list/}),
 and the Computer Aided CME Tracking (CACTus) catalog (\url{http://sidc.oma.be/cactus/}),
as well as STEREO Level 3 data, we confirmed that the solar activity was weak
between CIR1 and CIR2. During the period shown in Figure~\ref{fig:CIR_Observation},
 there were no $\sim >$A2.0 flares, no high frequency ($>$MHz) radio bursts, nor any fast
CMEs (the only detectable CME occurred on the east limb with a speed of 162 km s$^{-1}$ on DOY 239).
With these observations, any {\it in situ} enhancement of energetic particles is unlikely due to solar
activities and must be associated with the CIRs, as is consistent with the solar wind and magnetic
field signatures in Figure~\ref{fig:CIR_Observation}.

\begin{figure}[ht]
\includegraphics[width=0.9\textwidth]{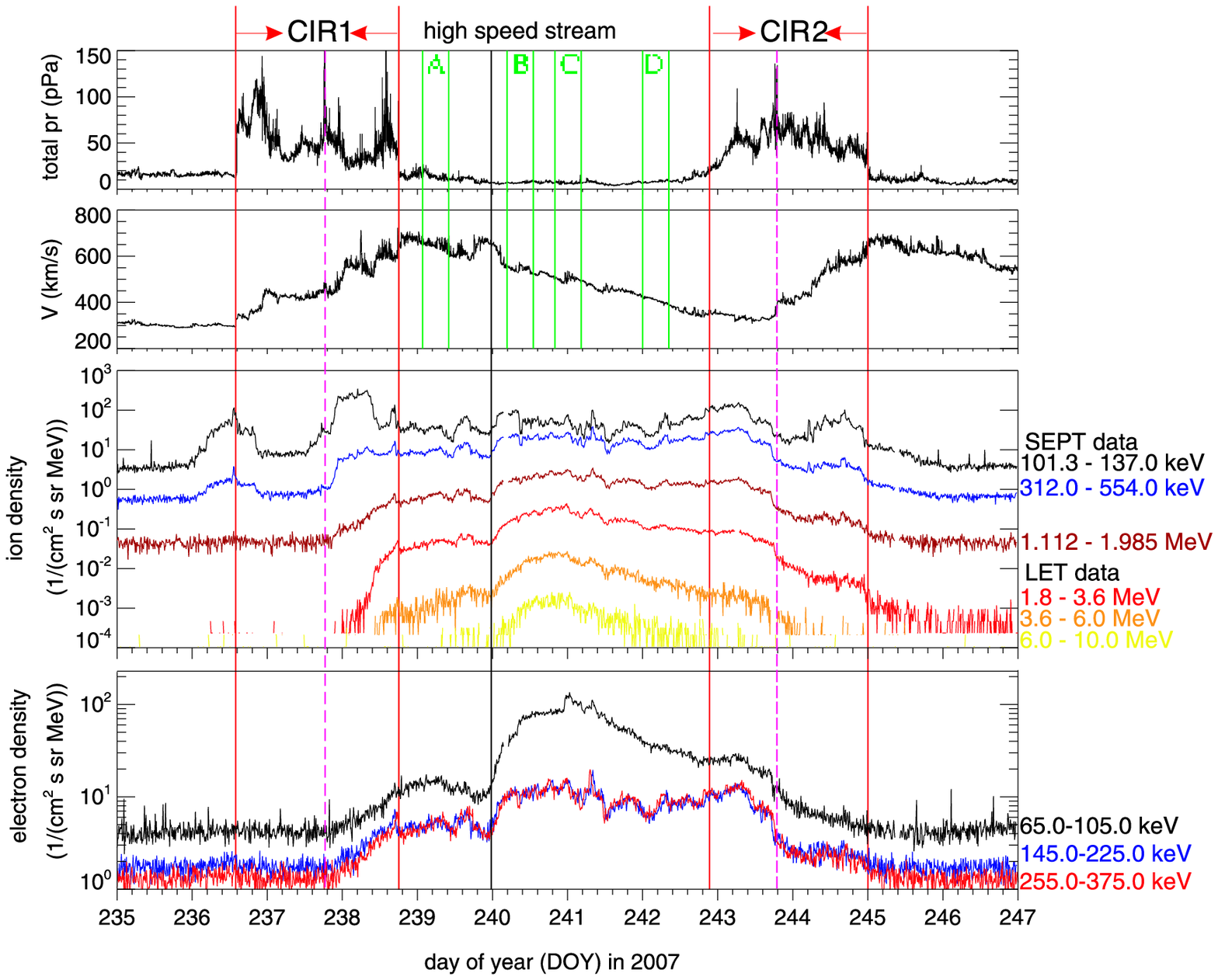}%{Fig4_SEP.eps}
\caption{STEREO-B observations of energetic electrons and ions of the CIR-pair event.
From top to bottom: solar wind total pressure, solar wind speed, ion and electron
time intensity profiles from IMPACT/SEP.
{ The four red vertical lines are the boundaries of the two CIRs and the two dashed vertical lines
are the stream-interface of the two CIRs. The black vertical line marks the occurrence of a
transient high-speed stream that occurred on DOY 240.} }
\label{fig:SEP-Observation}
\end{figure}

Figure~\ref{fig:SEP-Observation} shows fluxes of energetic particles observed by STEREO-B during the
period as shown in Figure~\ref{fig:CIR_Observation}. Panels 1 and 2 show the total pressure
(the thermal pressure plus the magnetic field pressure) and
the solar wind speed, as in Figure~\ref{fig:CIR_Observation}.
The third panel shows the ion (mostly protons) intensities for $6$ energy channels.
The fourth panel shows the electron intensity for three energy channels.
\blue{As in Figure~\ref{fig:CIR_Observation}, the boundaries of CIR1 and CIR2 are indicated by the
 solid vertical red lines. The stream interfaces within the two CIRs are shown as the dashed vertical
magenta lines. Four 8-hour intervals between the two CIRs, labeled ``A'', ``B'', ``C'', and ``D'' are
also indicated.
These periods will be used to examine the helium spectra and proton anisotropies
(see Figure~\ref{fig:SEP-Anisotropy} below).}

The third panel of Figure~\ref{fig:SEP-Observation} shows that there were gradual but significant
increases of ion fluxes in the $101.3$-$137.0$ keV and $312.0$-$554.0$ keV channels before the forward
shock of CIR1. The intensities reached their peaks at the
shock, then decayed to a plateau for a short period of time before they dropped at the beginning of day 237.
The plateau feature is typical of diffusive shock acceleration and has been seen previously in CIR events
(see eg. \citep{Bucik2009b}).
The drop inside CIR1 may be related to the increase in solar wind speed at the same time.
STEREO-B crossed the stream interface (left dashed line)
several hours before DOY 238, as shown by the left dashed line.
 Before the reverse shock of CIR1,
the intensities in the $1.1$-$2.0$ MeV  and  $1.8$-$3.6$ MeV channels showed clear gradual increases
which has also been seen in other CIR events \citep{Bucik2009b}. This is possible if the reverse shock was formed
inside 1 AU. Across the reverse shock, the intensities of both
the $1.1$-$2.0$ MeV and $1.8$-$3.6$ MeV ions slowly increased.
The $3.6$-$6.0$ MeV and $6.0$-$10.0$ MeV channels also showed clear increases.
Such increases contradict the scenario of diffusive shock acceleration at a single shock where
one expects to see a decrease of intensity as the distance from the shock increases (see e.g. the work of
\citet{Li.etal03, Li.etal05, Verkhoglyadova.etal10}).

The gradual increase of the ion time intensity profiles lasted until DOY 241. These increases were modulated by
a transient high-speed stream on DOY 240 (marked by the vertical black line).
The modulation showed no velocity dispersion and therefore could be due to
 connectivity issue, as discussed in \citet{Bucik2009b} for a different event.
The intensity peaks for all four channels occurred at the beginning of DOY 241. Beyond DOY 241, the intensities
for the two highest energy channels started to decrease.
Examination of Figure~\ref{fig:CIR_Observation} shows that $B_r$ reversed its sign on DOY 241.
\red{The large scale magnetic field configuration turns out to be an important key in understanding
this event. Below we examine the magnetic field configuration in detail, making use of the
suprathermal electron observations.}

At the leading edge of CIR2,
the four highest energy ion channels showed no increases. This is consistent with
the absence of a forward shock at CIR2. Across the stream interface (between DOY 243 and 244, as
indicated by the right dashed magenta line), all intensities dropped significantly.
Across the reverse shock of CIR2 at the beginning of DOY 245, the intensities showed further
gradual decreases, as one would expect from diffusive shock acceleration
in solar energetic particle events \citep{Li.etal03, Li.etal05, Verkhoglyadova.etal10}.

The fourth panel in Figure~\ref{fig:SEP-Observation} shows electron time intensity profiles for three energy
channels, $65.0$-$105.0$ keV, $145.0$-$225.0$ keV, and $255.0$-$375.0$ keV.
\blue{Similar to the ion observations, the most striking feature is the enhancement of intensity
 in between the two CIRs.
Indeed, for the $65.0$-$105.0$ channel, the peak of the time intensity profile occurred near the
middle of the two CIRs; for the $145.0$-$225.0$ keV and $255.0$-$375.0$ keV channels, the time intensity profiles
between the two CIRs are plateau-like. }
Such a peak and/or plateau-like feature has not been previously reported for single CIR events.

The time intensity profiles of ions and electrons between the two CIRs as observed by STEREO-B are very unusual for
CIR events.
Since the same CIR was also observed by ACE and STEREO-A, we next examine the
energetic particle observations at these spacecraft.

\begin{figure}[ht]
\includegraphics[width=0.9\textwidth]{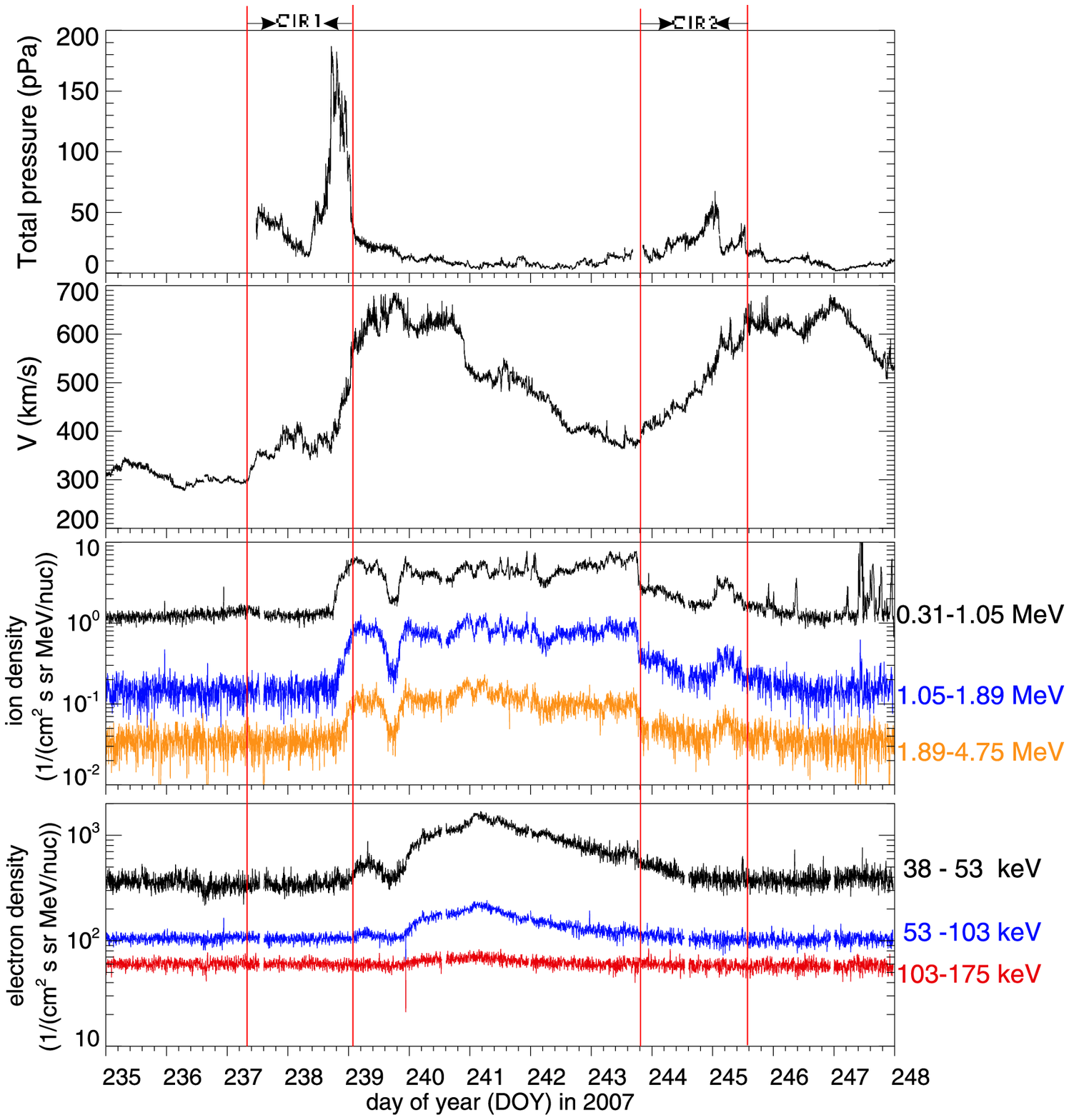}%{Fig5.eps}
\caption{ ACE observations of energetic electrons and ions of the CIR-pair event.
From top to bottom: solar wind total pressure,
solar wind speed, time intensity profiles  for ions (mostly protons) and electrons from ACE/EPAM.
{ The four vertical lines mark the boundaries of the two CIRs. }
}
\label{fig:SEP-Observation-ACE}
\end{figure}

\blue{ Figure~\ref{fig:SEP-Observation-ACE} shows the ACE observations.
The top panel shows the solar wind total pressure and the second panel shows the solar wind speed.
The lack of pressure at the beginning of the period is due to a data gap.
The third and fourth panels are ion and electron intensities from ACE/EPAM, respectively.
Comparing to the STEREO-B observations, while the $38$-$53$ keV and $53$-$103$ keV electrons show clear peaks,
the higher energy electrons (103-175 keV) and ions did not show clear peaks. The ion intensity
remains almost constant. }

\begin{figure}[ht]
\includegraphics[width=0.9\textwidth]{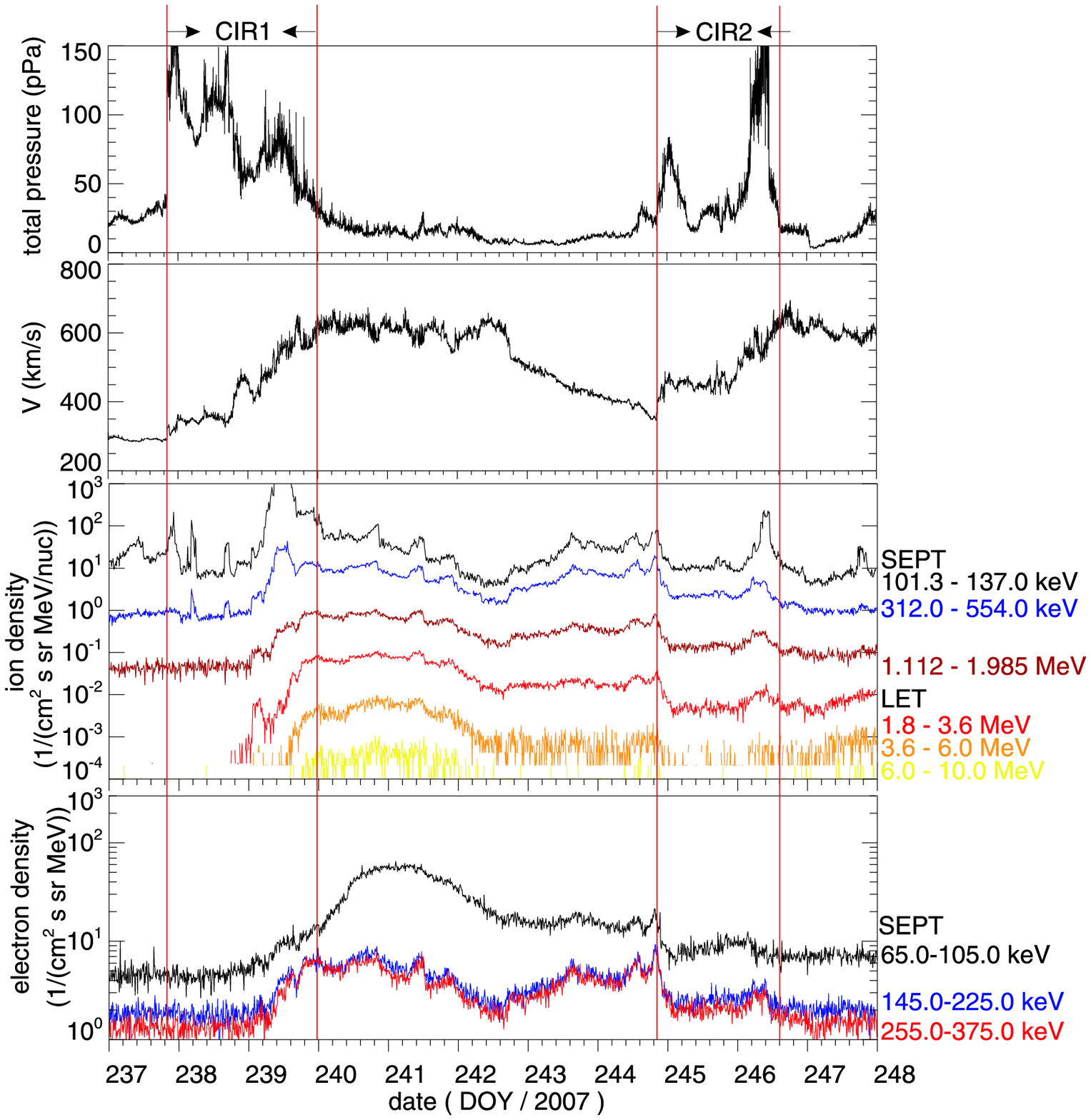} %{Fig6.eps}
\caption{STEREO-A observations for the CIR-pair event.
From top to bottom:  solar wind total pressure, solar wind speed,
time intensity profiles from IMPACT/SEP ion and electron measurements.
{ The four solid vertical lines are the boundaries of the two CIRs.} }
\label{fig:SEP-Observation-STA}
\end{figure}

\blue{Figure~\ref{fig:SEP-Observation-STA} shows the STEREO-A observations.
As in Figure~\ref{fig:SEP-Observation-ACE}, the top two panels display
the total solar wind pressure and
the solar wind speed, respectively.
The third panel shows the ion (mostly proton) intensities from SEPT and proton intensities from LET,
while the fourth panel shows the electron intensities from LET.
Again, there were no clear peaks in the low energy ions between the two CIRs.
However, for high energy ions observed by LET (especially $3.6$-$6.0$ MeV and $6.0$-$10.0$ MeV channels),
intensities gradually increased after CIR1, until the middle of DOY 241, and
then quickly dropped. Compared to ions, electrons showed more pronounced peaks after CIR1.}

\red{SEP observations at ACE and STEREO-A showed similar unusual behavior,
 but less prominent. This is possible if the underlying cause for the STEREO-B observation is transient in
nature and affects ACE and STEREO-A observation for a shorter period of time.
 Indeed, this is an important clue for understanding this event, as discussed later.}

\begin{figure}[ht]
\includegraphics[width=0.9\textwidth]{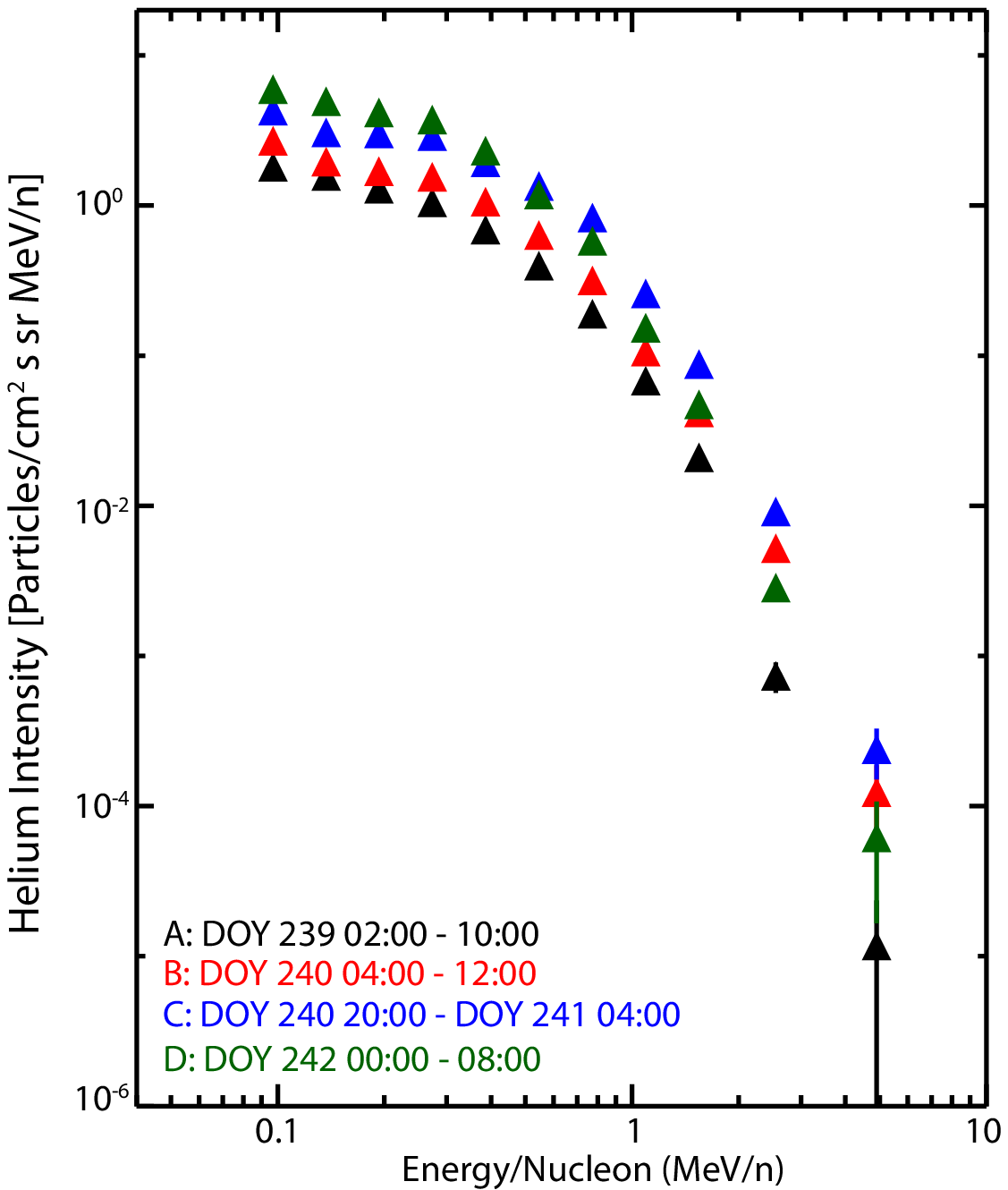}%{Fig7_spec.eps}
\caption{ The spectra of energetic helium in the four periods: ``A'', ``B'', ``C'', and ``D'' shown in
 Figure~\ref{fig:SEP-Observation}.}
\label{fig:spectra}
\end{figure}

We now examine how the spectrum of energetic helium ions
evolved in this event. Figure~\ref{fig:spectra} shows the energetic
helium spectra during the four 8-hour
intervals, ``A'', ``B'', ``C'', and ``D'' shown in Figure~\ref{fig:SEP-Observation}.
For example,  at $\sim 0.1 $MeV/nucleon, the helium intensity during
period B (C) is about twice as large as in period A. At $\sim 2 $MeV/nucleon,
the helium intensity during periods B and C is more than $10$ times larger than in period A.
The spectrum for period D is similar to the spectrum for period A, but the intensity was higher.
These spectral observations show that there are more high energy particles near the middle of the two
shocks (periods B and C) than closer to the shocks (periods A and D).  Note that
period A is closer to the reverse shock of CIR1 and period D is closer to the  forward shock of CIR2.

The four panels in Figure~\ref{fig:SEP-Anisotropy} show sector plots of the anisotropy of $6$ to $10$ MeV
 protons in the RT plane for the four 8-hour intervals, respectively.
The Radial, Tangential, Normal (RTN) coordinate system is used.
The $R$ direction is from the Sun to the spacecraft and is pointing to the left;
the $T$ direction is along the $\Omega \times \hat{R}$  direction, where $\Omega$ is the Sun's
spin axis; and the $N$ direction completes the right-handed triad.
The Sun is to the right in these plots.
\blue{At $1$ AU, the nominal magnetic field with a Parker spiral
orientation (pointing away from the Sun) will be along
the diagonal from the lower right corner to the upper left corner (the arrow shown in panel A).
We use data from the LET instrument, which has a total of $16$ sectors with $8$ facing toward
and $8$ against the Sun. The convention of the plot is such that the upper left (lower right) sector
measures ions coming from the upper left (lower right). The lengths of the sectors denote the
intensities of the fluxes as seen by that detector with the scale shown in the $x$ and $y$-axis.
The red and blue lines shown in the panel represent the $2$-minute averaged magnetic field directions,
with red (blue) indicating negative $B_r$ and blue positive $B_r$.}

In panel A, except for two intervals, the magnetic field points toward the Sun
and clustered around the Parker spiral direction, but with a large spread.
The proton flux was lower than those in the other three panels. The flux was rather isotropic,
with slightly more particles coming into the $8$ upper-left detectors than coming into the $8$ lower-right
detectors.

In panel B, the directions of the magnetic field still points toward the Sun.
The spread of the magnetic field direction is narrower than in panel A and the center of the distribution
is slightly inclined towards the R direction. The flux is more than $10$ times higher
than those in Panel A and there is a clear anisotropy in that more protons are propagating towards
the Sun than away. The anisotropy suggests that the source of the $4$-$6$ MeV protons was located
outside 1 AU. These protons were presumably accelerated at the reverse shock of CIR1 (now beyond 1 AU)
and propagated back to $1$ AU. Note that the fluxes detected by the $8$ detectors facing against the Sun
(the upper left $8$ sectors) do not differ much between themselves,  $\sim 0.012$/(cm$^2$ s sr MeV/n).
Similarly the fluxes detected by the $8$ detectors facing the Sun (the lower right $8$ sectors) did
 not differ much, $\sim 0.007$/(cm$^2$ s sr MeV/n).
So the fluxes are isotropic in each hemispheres.
Note that particles propagating towards the Sun can be reflected, and then as they move back away
from the Sun,
their pitch angles will approach zero due to the ``magnetic focusing'' effect.
However, the fact that the pitch angle distributions (PADs) were isotropic within each
hemisphere suggests that pitch angle diffusion
dominated over magnetic focusing. As a measure of the anisotropy, we find the ratio of the
sunward to anti-sunward fluxes is  $\sim 2$. This value
depends very much on the pitch angle diffusion coefficient
and can be used to constrain models of particle acceleration and transport in CIRs.

Panel C is for the period where the peak of the $6$-$10$ MeV proton intensity occurred.
The magnetic field showed a clear bi-polar distribution during this period:
when $B_r < 0$ (the red lines), the magnetic fields were generally aligned with the nominal Parker
 spiral direction;
when $B_r > 0$ (the blue lines), the magnetic fields were aligned mostly
between the radial direction and that is perpendicular to the nominal Parker spiral direction.
Compared to panel B, the proton flux was enhanced. The pitch angle distribution was almost isotropic,
with slightly more particles propagating away from the Sun.

The period shown in panel D was closer to CIR2.
During this period $B_r>0$, so the magnetic field was pointing away from the Sun.
Period D was thus on the opposite side of the HCS.
The proton flux was typically smaller than
in periods B and C (note the scales in panel A and D are different
from those in panels B and C), but higher than those in period A.
There was a significant anisotropy in the angular distribution, with more protons
propagating away from the Sun.

\begin{figure}[ht]
\includegraphics[width=0.95\textwidth]{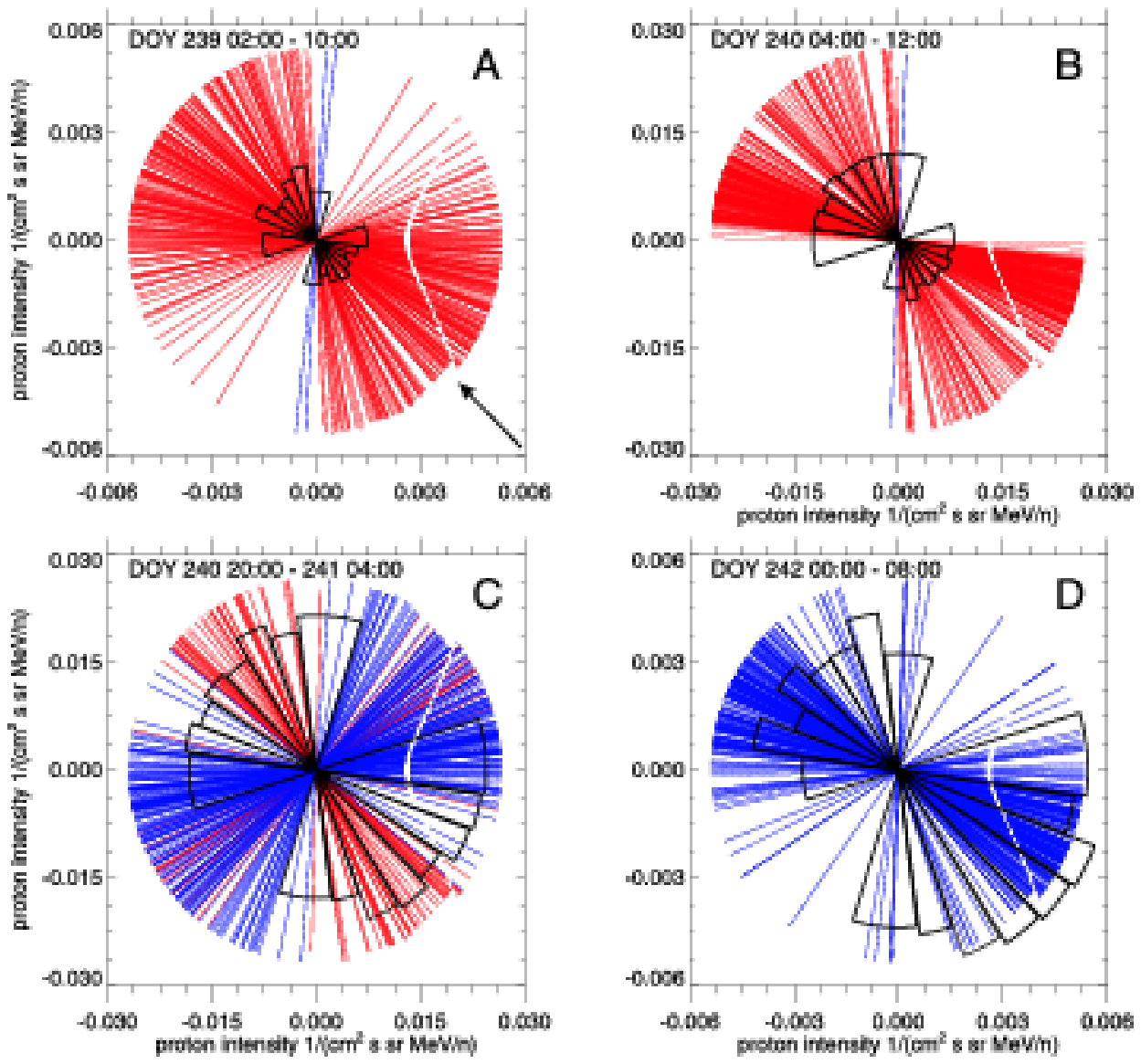}%{Fig8_Anisotropy.eps}
\caption{The anisotropy plots of $6$-$10$ MeV protons from the STEREO-B LET instrument
during the intervals
A-D in Figure~\ref{fig:SEP-Observation}.
The four panels are the sector plots of protons; note different intensity scales for the panels.
 The RTN coordinate system is used. The Sun is to the right.
The length of the sectors represent the intensity of protons in the $16$ sectors, respectively.
Red and blue lines show 1 minute average IMF directions, with red indicating negative $B_r$, blue
indicating positive $B_r$. }
\label{fig:SEP-Anisotropy}
\end{figure}

\red{
The observations of energetic ions and electrons described above are very unusual and were never reported
in individual CIR events.
For example, in typical single CIRs the maximum intensities occur near either the
forward or the reverse shock, and after spacecraft passage
through the reverse shock, the intensities usually drop quickly.
Furthermore the anisotropy of SEPs after the reverse shock is usually sunward. }

\section{Discussion}
\red{What caused all these unusual observations in this event?
The presence of the second CIR likely plays an important role.
However these observations are also rare for CIR pairs. Indeed as we will discuss in the following,
for the Carrington Rotation
 immediately before and after CR2060, where CIR pairs also appeared,  energetic ion intensity
observations showed the normal temporal behavior of single CIR cases. This is somewhat expected because
unless the two CIRs are magnetically connected, the
presence of a second CIR will have no effects on the energetic particles associated with the first CIR, and
vice versa. }

\begin{figure}[ht]
\includegraphics[width=0.9\textwidth]{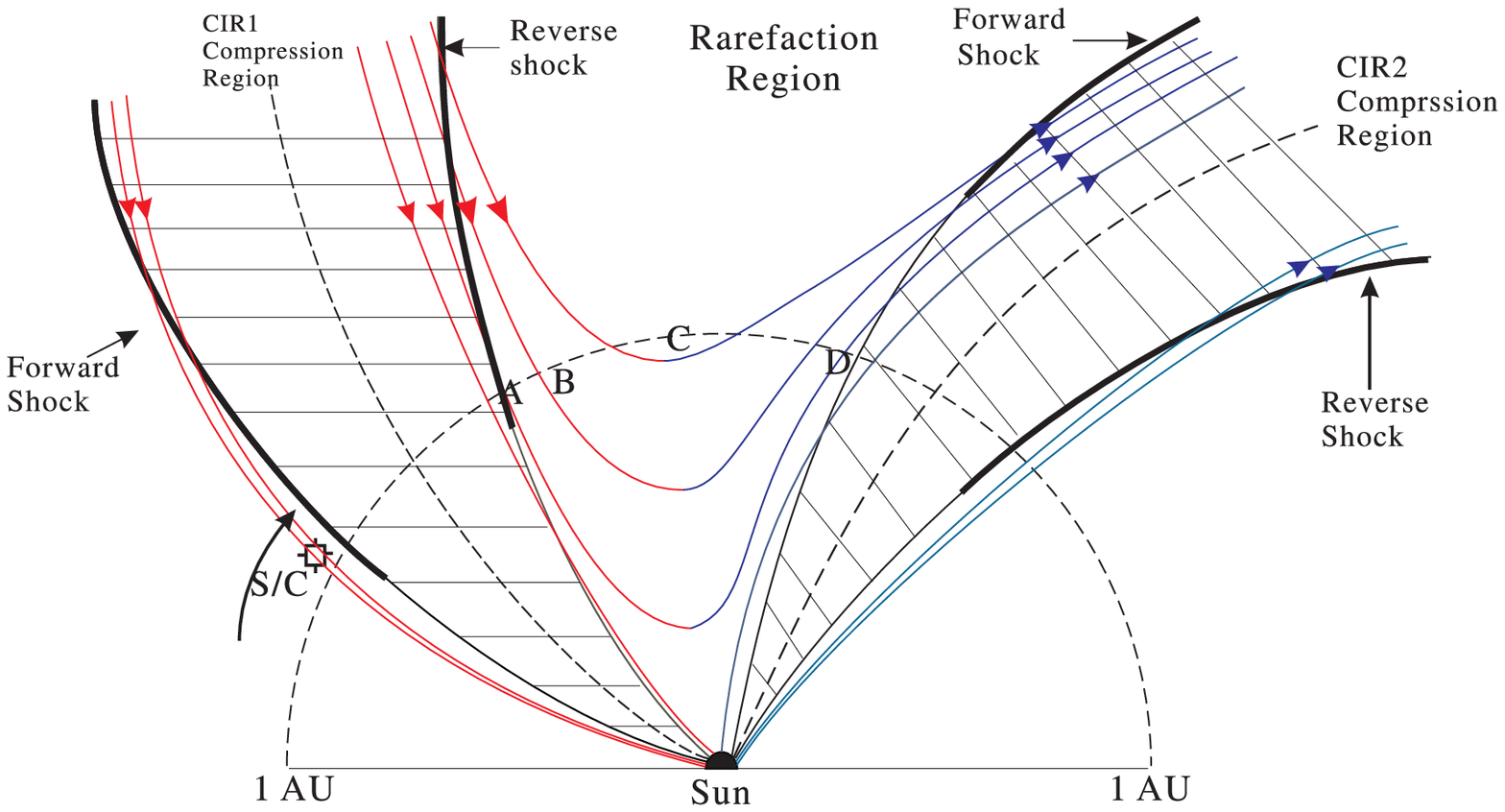}%{Fig9_Cartoon.eps}
\caption{Cartoon showing the configuration of the magnetic field in the CIR-pair event. We conjecture that
the U-shaped magnetic field is due to magnetic reconnection that occurred at high corona.
The color shows the sense of the magnetic field: red is negative (arrows pointing to the Sun) and blue is positive
(arrows pointing away from the Sun).
 The spacecraft trajectory, as seen from the co-rotating frame, is clockwise along the semi-circle. The two shaded areas
are the compression regions. The region between the two CIRs is the rarefaction region. The forward and the
reverse shocks are shown as the dark black curves. Note that as shown in the cartoon,
the forward shock and the reverse shock of CIR1 and the reverse shock of CIR2 are formed
inside 1 AU, while the forward shock of CIR2 is formed outside 1 AU. The U-shaped magnetic field line is proposed
to account for the observations of both the reversal of the magnetic field sense
and the anisotropy observations of energetic particles between the two CIRs.}
\label{fig:Cartoon}
\end{figure}

\red{The above argument suggests that magnetic connection is important in the case of CR2060.
We argue in the following that these observations (as shown in Figures~\ref{fig:SEP-Observation}
to~\ref{fig:SEP-Anisotropy}) resulted from particle acceleration and trapping between the reverse shock of
the first CIR and the forward shock of the second CIR through a large scale ``U-shape'' magnetic field that
connects these two shocks. We illustrated schematically in Figure~\ref{fig:Cartoon}.
The U-shaped magnetic field was possibly caused by reconnection that occurred high in the corona.
Previously, \citet{Luhmann1998} and \citet{Li2001} suggested that such reconnections may
constantly occur near the cusp of the helmet steamers
 due to the continuous global evolution of coronal fields.
In a numerical modeling effort, \citet{Chen.etal09} noted that reconnections can take place
above a streamer cusp repeatedly as a result of insufficient
magnetic confinement of streamer plasmas. }

In Figure~\ref{fig:Cartoon},
the dashed semicircle is the approximate spacecraft trajectory with respect to the Sun in the co-rotating
frame. The spacecraft moves along the semicircle in the clockwise direction (as shown by the arrow).
The labels A, B, C, D correspond to the 4 periods shown in Figure~\ref{fig:SEP-Observation}.
The two compression regions are indicated by the horizontal and inclined lines.
The stream interfaces are shown as the dashed lines inside the two compression regions.
The solid lines in red and blue and with arrowheads are the magnetic
field lines. Red denotes a negative $B_r$ and blue denotes a positive $B_r$.
The four dark black lines are the four shocks. The forward shocks are formed in the slow wind and the reverse shocks
are formed in the fast wind. The forward and reverse shocks bounded the CIRs. Note that as drawn in
Figure~\ref{fig:Cartoon}, both the forward and reverse shocks of CIR1 are formed inside 1 AU.
In comparison, the forward shock of CIR2 is formed outside 1 AU, but the reverse shock is formed inside 1 AU,
as required by the {\it in situ} observations as shown in Figure~\ref{fig:CIR_Observation}.
Three U-shaped magnetic field lines are shown in Figure~\ref{fig:Cartoon}.
These are drawn to help understanding the time intensity profiles shown in
Figure~\ref{fig:SEP-Observation} and the anisotropy observations shown in Figure~\ref{fig:SEP-Anisotropy}.

Referring to features in Figure~\ref{fig:Cartoon}, we now describe a scenario to explain the STEREO-B observations.
 Consider first period A, at this time STB had just passed the newly developed reverse shock from left (downstream)
to right (upstream). It was close to the shock, so the distribution of the energetic particles did not show a
strong anisotropy. Furthermore, since the shock was newly formed at this distance, the acceleration was  less
efficient and the intensity of the energetic particles was low.
As STB moved clockwise to period B, it connected to the part of the shock that was further away, where the shock was
stronger. Since the shock was the source of the energetic particles, we expect to see most energetic particles
propagating sunward during this period, as observed in Figure~\ref{fig:SEP-Anisotropy}.
Although most particles were flowing toward the Sun,
$\sim 1/3$ were moving away from the Sun.
This is not unusual in CIR events where the pitch angle anisotropies
are usually modest.
A small anisotropy is likely due to a strong pitch angle scattering.
Furthermore magnetic reflection of sunward flowing particles, as they move closer to the Sun, can
reverse their propagation directions.
So even if pitch angle scattering is not strong, these reflected particles will be observed as moving away from the Sun.

Moving from period A to B, the particle intensity increases, and the anisotropy also increases. In
the U-shaped magnetic field scenario, this is due to the spacecraft being connected to stronger sources (further
away). As the spacecraft further moved to period C, we find that the direction of the magnetic field went through
a phase during which it was mostly perpendicular to the nominal Parker field direction and eventually
the $B_r$ flipped sign.
This can be explained naturally with the U-shaped magnetic field, since in this case,
the B field direction had to undergo a change from a Parker-spiral-like with $B_r<0$ to an intermediate state
where $B_r=0$ and finally to a direction with $B_r>0$.
During period C, the intensity of energetic ions also increases. This is not surprising since now the
U-shaped field lines connected to both the reverse shock of CIR1 and the forward shock of CIR2, so that accelerated
particles are effectively trapped between the two shocks.
Furthermore, since it was in the middle of CIR1 and CIR2, it connected to the furthest and strongest parts
of both shocks, consequently the acceleration was the most efficient.
Also note that further away the shock compression ratio could be larger and the particle spectrum can be therefore harder.
That particles are now trapped between two shocks also implies
that the anisotropy in period C will be smaller than that in period B.
This is what the observations show in Figure~\ref{fig:SEP-Anisotropy}.
We note that for energetic electrons, because they move fast,  they may undergo multiple reflections between the
reverse shock of CIR1 and the forward shock of CIR2. This may lead to a more efficient acceleration than that at a
single shock.

In period D, the U-shaped field lines were still connected to the reverse shock of CIR1, but not connected to
the forward shock of CIR2 because it was not yet formed at 1 AU
(during period D the field lines are the same as those in period A and B).
The U-shaped magnetic field is not connected to the forward shock of the CIR2, so only limited
reflection can occur locally in CIR2, and
therefore the energetic particles in period D were mostly streaming away from the Sun.
This agrees with the anisotropy observations in Figure~\ref{fig:SEP-Anisotropy}. Furthermore, the absence of the
reflection also implies that the intensity of energetic ions in period D
will drop comparing to period C. This agrees with the time-intensity profile
observations shown in Figure~\ref{fig:SEP-Observation} and
Figure~\ref{fig:SEP-Anisotropy}. Note that the intensity in period D is comparable to those
in period A and B. Also we remark that magnetic field
enhancement in a CIR without shock can also reflect particles, but presumably
of less efficiency than if there is a shock.

\red{ In our proposed scenario, the large scale U-shaped magnetic field topology is the key. So
it is important to see if there is corroborating observational evidence for this.
As explained in the introduction, solar wind suprathermal electrons can be used to infer the
large-scale topology of the interplanetary magnetic field (e.g., \citet{Gosling1990, Kahler1994}).
Recent studies \citep{Steinberg2005, Skoug2006, Lavraud2010} showed that counter-streaming
suprathermal electrons are frequently observed in the vicinity of CIRs,
owing to wave-particle interactions and shock heating, combined with adiabatic mirroring
and particle leakage into the upstream regions (both into the adjacent slow and fast winds).}

\begin{figure}[ht]
\includegraphics[width=0.9\textwidth]{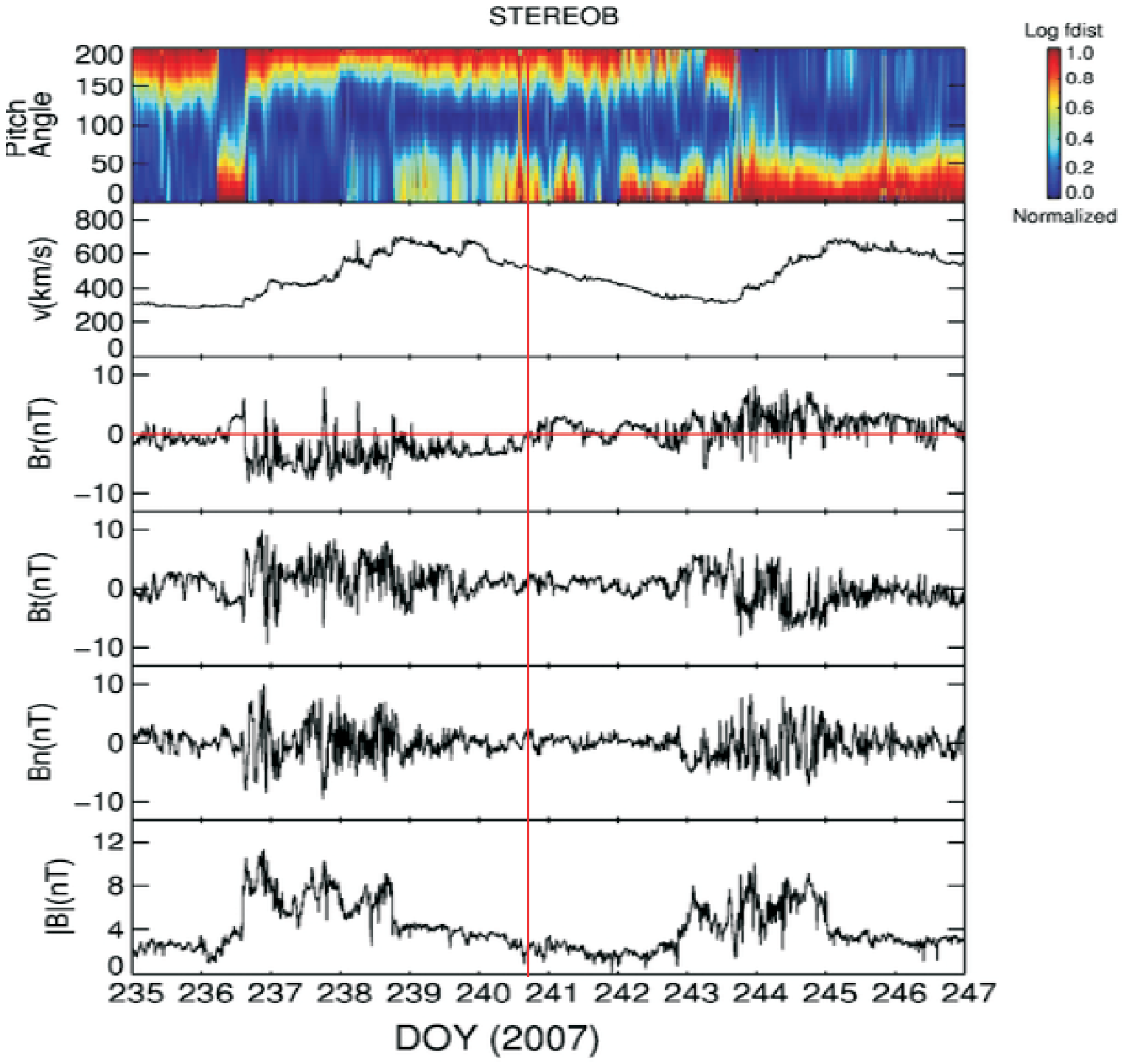}%{STB_B_Strahl.pdf}
\caption{STEREO-B observations from top to bottom: the PAD of $246.62$ eV strahl electrons;
solar wind proton speed;
$R$, $T$, and $N$ components of the vector magnetic field; and total magnetic field magnitude.
The reversal of $B_r$ is indicated by the vertical red line. }
\label{fig:STB_B_Strahl}
\end{figure}

\red{However, statistical analyses \citep{Lavraud2010} showed that such counter-streaming features
are only observed to some limited distance downstream of the CIR bounding shocks/compressions.
In other words, they are not expected to extend all the way from one CIR to the next.
We thus now examine the pitch angle distributions (PAD) of $\sim 250$ eV electrons in the context of the
proposed U-shaped topology. We first consider STEREO-B observations. In Figure~\ref{fig:STB_B_Strahl},
the top panel displays the normalized (for each sample) electron PAD, the second panel shows the solar wind
speed, and the next four panels show the magnetic field components and its magnitude. This figure shows a dominant
anti-field-aligned strahl until DOY $\sim$239. }

\red{On day 236, we note a short duration reversal of the dominant PAD polarity. However, this is commonly
observed when strong acceleration and reflection of suprathermal electrons occurs at CIR shocks
(cf. \citep{Steinberg2005, Skoug2006}). Although not shown, the fluxes of the $180^\circ$ population
actually remained the same as adjacent regions during this short interval, but the overwhelming fluxes of the
reflected population make the normalized PADs appear dominated by this reflected population.
}

\red{The most important PAD feature is seen between the two CIRs. Starting slightly before DOY 379 at the time of
the reverse shock of CIR1, and lasting until the beginning of CIR2 (DOY 243), STEREO-B recorded unusual
and essentially continuous counter-streaming PADs. As shown in previous studies, this feature is not
necessarily related to a closed field topology. It is often seen upstream of CIR shocks, but again what
makes it unusual here is that it is observed continuously between the two CIRS. In the proposed
scenario of a U-shaped magnetic topology connecting the reverse shock of CIR1 and the forward shock of CIR2,
counter-streaming suprathermal electrons is a natural consequence; for the same reasons as explained previously
for more energetic ions and electrons. A U-shaped magnetic topology also implies that $B_r$ will reverse sign,
as indicated by the red vertical line in Figure 9. Note that the $B_r$ reversal is in any case
expected between two consecutive CIRs, if these do not come from pseudo streamers, as is the case here.}

\begin{figure}[ht]
\includegraphics[width=1.0\textwidth]{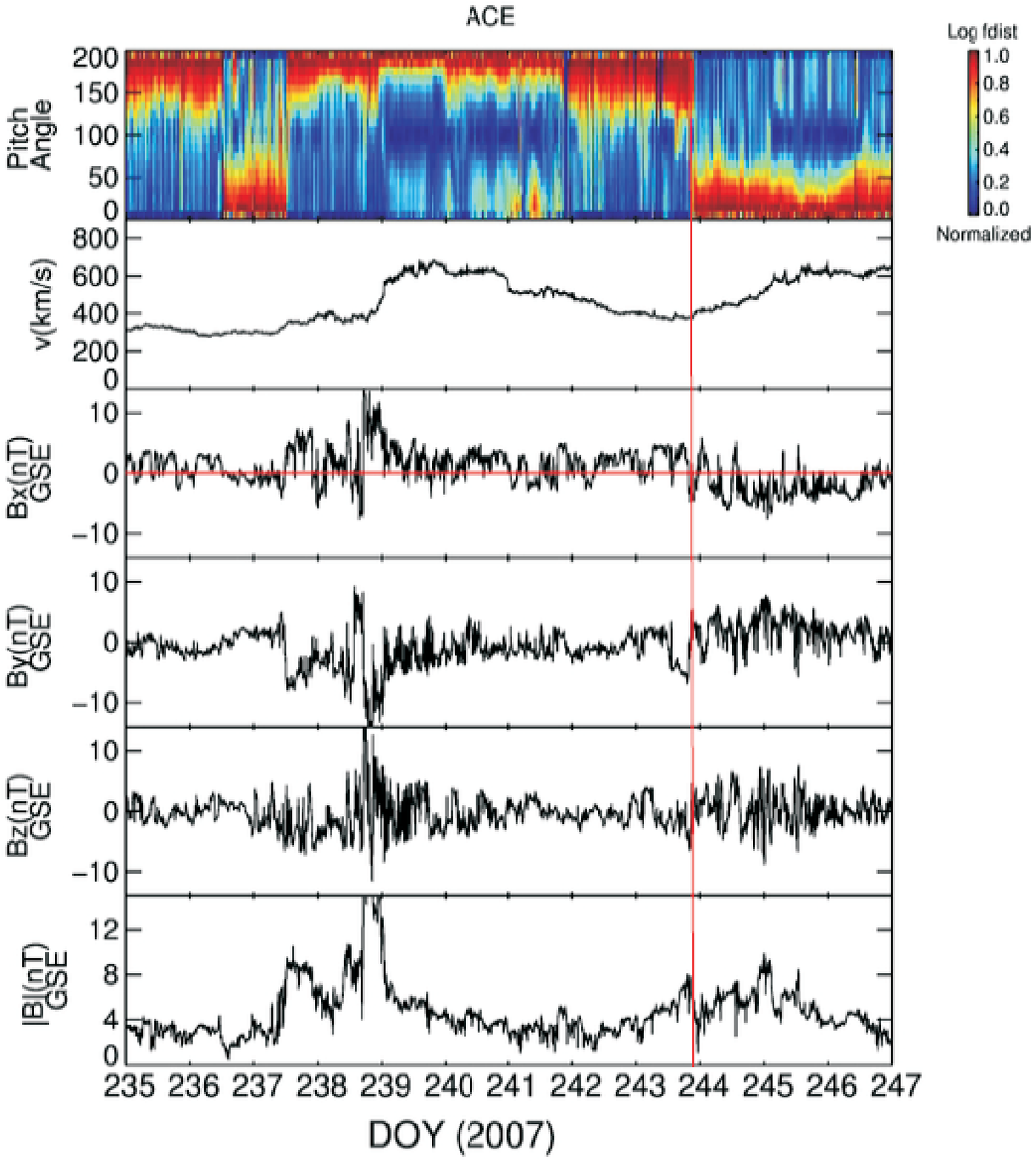}%{ACE_B_Strahl.pdf}
\caption{Similar to figure~\ref{fig:STB_B_Strahl}, but for ACE. Note that the ACE observation uses the GSE coordinate
system, while STEREO-A and B use the RTN system. The reversal of $B_r$ is indicated by the vertical red line. }
\label{fig:ACE_B_Strahl}
\end{figure}

\begin{figure}[ht]
\includegraphics[width=1.0\textwidth]{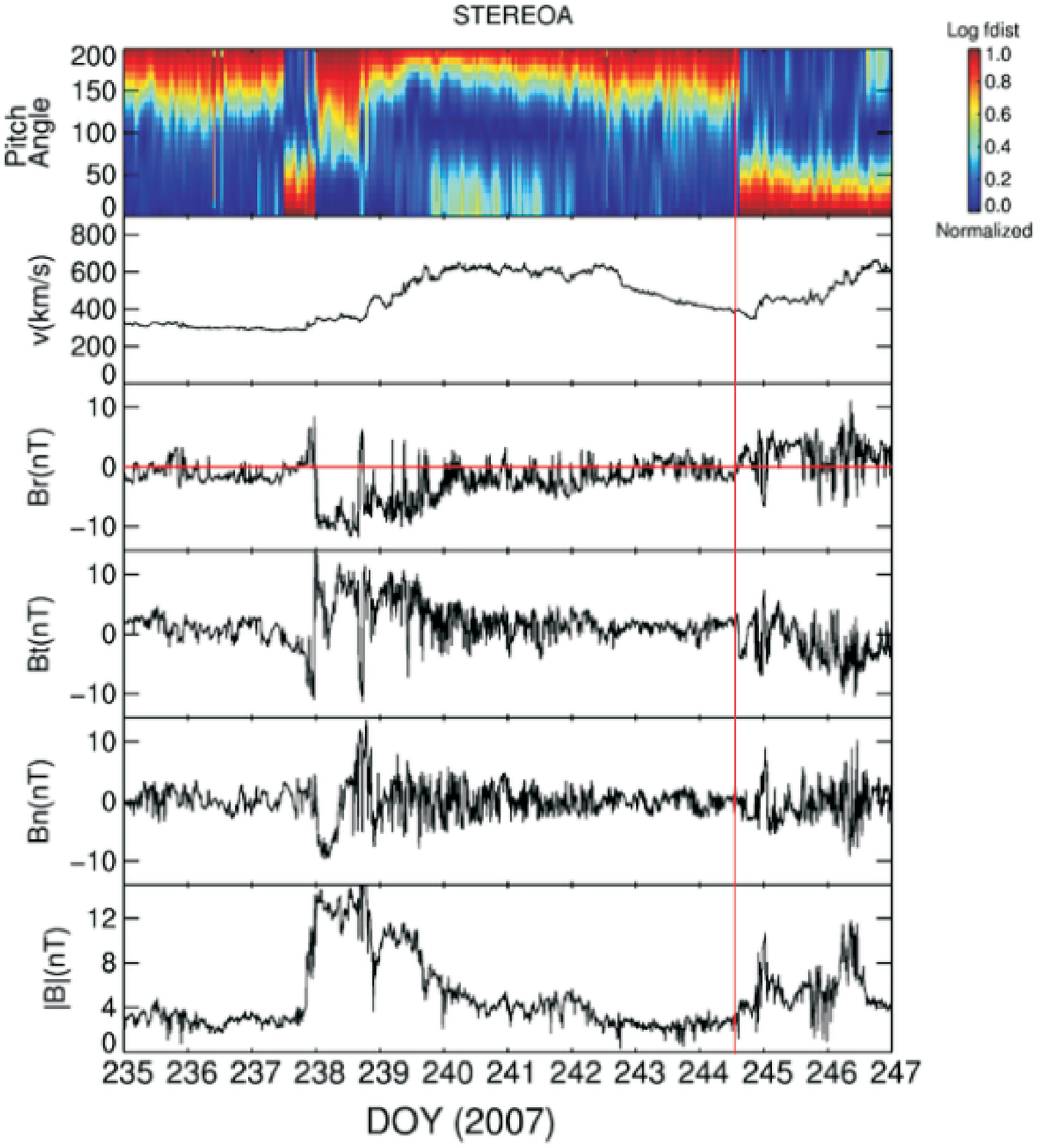}%{STA_B_Strahl.pdf}
\caption{Similar to figure~\ref{fig:STB_B_Strahl}, but for STEREO-A. Note that the ACE observation uses the GSE coordinate
system, while STEREO-A and B use the RTN system. The reversal of $B_r$ is indicated by the vertical red line. }
\label{fig:STA_B_Strahl}
\end{figure}

\red{Figure~\ref{fig:ACE_B_Strahl} and Figure~\ref{fig:STA_B_Strahl}  are similar to Figure~\ref{fig:STB_B_Strahl},
but for ACE and STEREO-A, respectively.
Compared to STEREO-B, the counter-streaming features at ACE and STEREO-A are much less prominent,
and of shorter durations. The vertical red lines mark the times when $B_r$ changes sign at ACE and STEREO-A.
In both cases, $B_r$ reversed sign in the slow solar wind and close to CIR2, as is usually the case. This is
unlike the $B_r$ sign change at STEREO-B which is very significantly farther from CIR2. We take these observations
as additional arguments in favor of a large-scale U-shaped topology, as depicted in Figure 8, which is caused
by magnetic reconnection in the high corona. This phenomenon is transient in nature since ACE and STEREO-A measurements
do not show the same continuous PAD properties as STEREO-B. }

As discussed next, these properties are also not found for the previous and next Carrington rotations.

\subsection{Energetic particle observations from the preceding and the following Carrington Rotations} \label{sec:Discussion}

\begin{figure}[ht]
\includegraphics[width=0.9\textwidth]{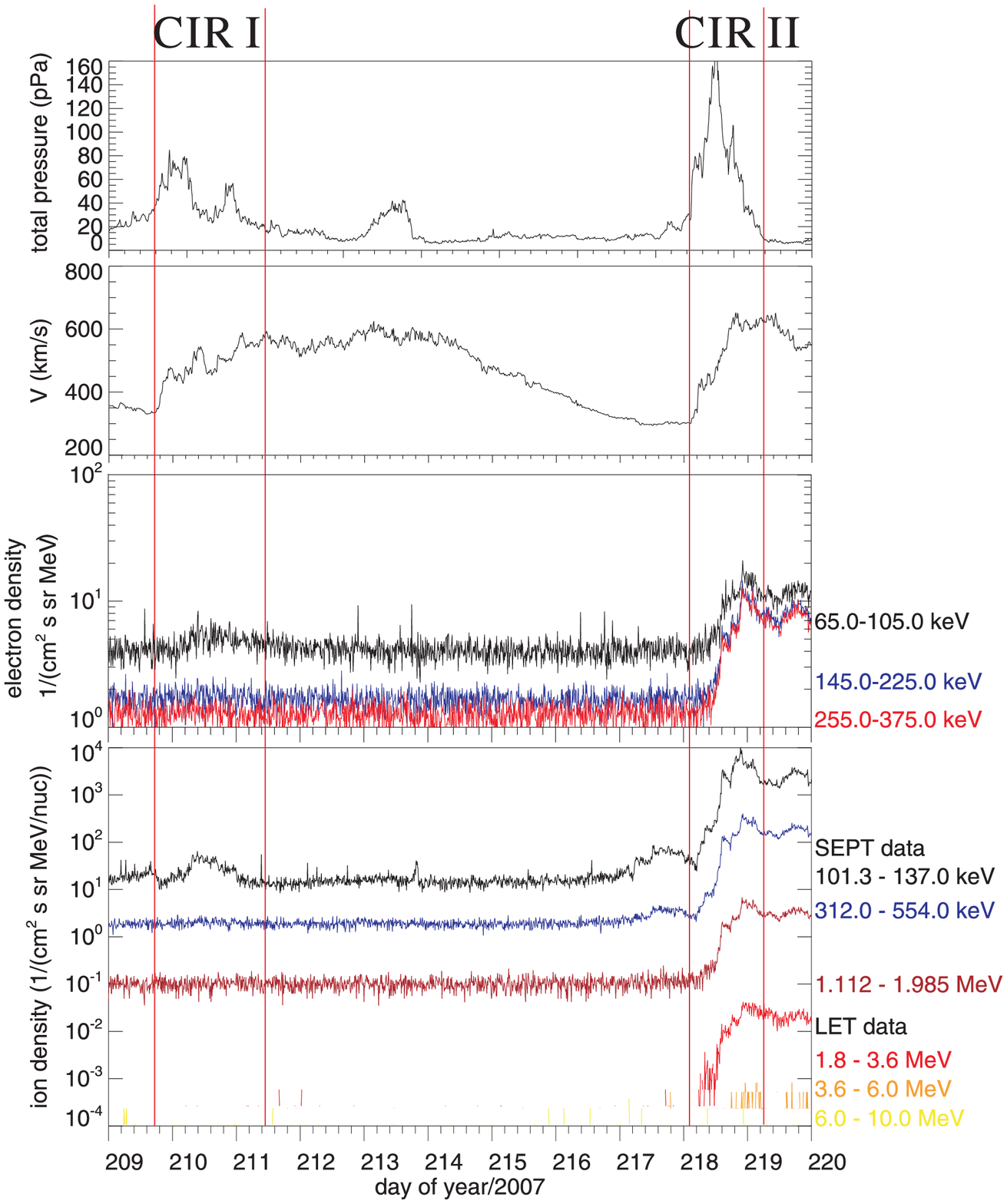}%{Before_event.eps}
\caption{CIR pair observation of CR2059. From top to bottom:
plasma total pressure, solar wind speed, intensities for
energetic electrons (3 energy channels), and intensities for
energetic proton and ions (6 energy channels). }
\label{fig:preCIR}
\end{figure}

\begin{figure}[ht]
\includegraphics[width=0.9\textwidth]{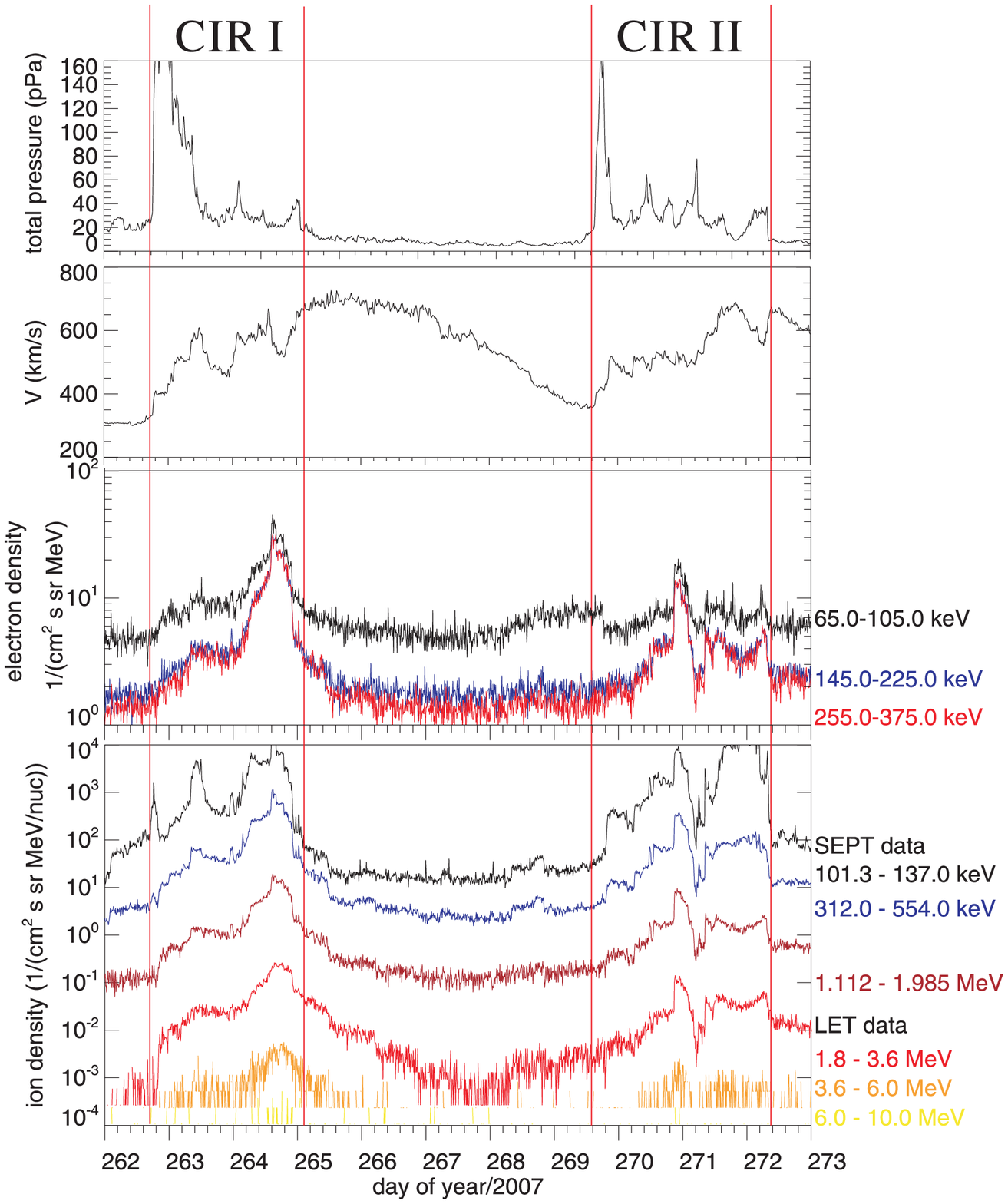}%{After_event.eps}
\caption{Same as Figure~\ref{fig:preCIR}, but for CR2061.}
\label{fig:aftCIR}
\end{figure}

\red{The observations discussed so far have motivated us to conjecture the existence of a large scale U-shaped magnetic
field topology between the reverse shock of CIR1 and the forward shock of CIR2.
The U-shaped magnetic field was  possibly caused by reconnection that occurred high in the corona. As such, it
is expected to be transient in nature, which explains why these unusual energetic particle
 observations were most prominent at one spacecraft, in this case STEREO-B.
Such a transient nature of the U-shaped magnetic field also implies that
similar observations are not expected to emerge for the CIR pairs before and after CR2060.
We now examine the energetic particle observations from the CIRs
in the preceding and following Carrington Rotations by STEREO-B. These are shown
in Figure~\ref{fig:preCIR} and Figure~\ref{fig:aftCIR}.}

\red{For CR2059, while the two high speed solar wind streams are evident, the shock pair that bounded the
first CIR could not be identified. In comparison, the forward shock, but not the reverse shock, of the second CIR has
developed at 1 AU.  There was clear signatures of particle acceleration at the second CIR as shown by the
enhancement of electrons and ions in most of the energy channels.
The intensities of the $65$ to $105$ keV electron and $101.3$ to $137.0$ keV/nucleon ion showed noticeable enhancement
at the first CIR. The time intensity profiles between the two CIRs showed a very regular behavior as expected since there
was no reverse shock associated with the first CIR.}

\red{For the CR2061, the shock pairs that bounded both CIRs can be clearly seen, making it a good comparison to CR2060.
Compared to Figure~\ref{fig:SEP-Observation}, we see stark differences in the
time intensity profiles between the CIR pairs.  In CR2061, there were the normal drop-outs followed by
gradual enhancements of energetic electrons and ions between the two CIRs. This is best illustrated by the $1.8$-$3.6$ MeV
ion channel. Without the U-shaped magnetic field topology,
 this drop-out is a normal behavior since the spacecraft moved further away  from the reverse shock of the first CIR.
Similarly the subsequent gradual increase is due to the spacecraft becoming closer to the forward shock of the second CIR. }

\red{So the presence of a CIR pair is only a necessary condition for the observations shown in
Figure~\ref{fig:SEP-Observation}. In our proposed scenario, the essential requirement for the unusual intensity
enhancements between the two CIRs is a field that connects the two CIRs (to be more specific,
the reverse shock of the first CIR and the forward shock of the second CIR).  This is achieved through magnetic
reconnection in the high corona.}

\section{Conclusion}
We report the observation of the acceleration and trapping of energetic ions and electrons
at a CIR pair. The event occurred in Carrington Rotation (CR) 2060 and the two CIRs,
with a separation less than $5$ days, were observed by three spacecraft: STEREO-B, ACE and STEREO-A.

Several distinct observational features in this CIR-pair event made this event very unusual.
These features are:

1) The time intensity profiles of high energy ions and electrons show a  peak between the two CIRs.

2) The magnetic field changed sign in between the CIR-pair, in a  fashion consistent with regular CIRs, rather
 than CIRs from a pseudostreamer.

3) The PAD of high energy protons between the two CIRs showed an unusual sequence of
first being nearly isotropic, followed by mostly sunward propagating, then isotropic, and
finally mostly anti-sunward propagating.

4) The energetic helium spectra  were harder in between the two CIRs than near the CIR shocks.

5) Counter-streaming suprathermal electrons were observed continuously between CIR1 and CIR2.

\red{These features have motivated us to propose a  U-shaped large-scale magnetic field topology
connecting the reverse shock of the first CIR and the forward shock of the second CIR.
A  U-shaped magnetic field can simultaneously explain all the above observations.
The key point is that a magnetic field connecting the two shocks
(the reverse shock of CIR1 and the forward shock of CIR2) will allow particles to be
 trapped and accelerated between the two shocks.
We conjecture that magnetic reconnection occurring in the high corona is the cause of such a
U-shaped magnetic field. After reconnection, the U-shaped magnetic field propagates out to the heliosphere.
 We infer that the magnetic reconnection process is transient in structure because observations
from ACE and STEREO-A for the same event were less prominent and because there were no similar observations
during the preceding and following Carrington rotations.
In a future paper, we plan to examine particle acceleration at a CIR-pair with a U-shaped magnetic field
configuration using a test particle simulation. }

\acknowledgments
This work is supported in part by NSF grants AGS1135432 and ATM0847719, and NASA grant NNX13AE07G at UAH,
NNSFC41331068, 41274175, 41028004, and NSBRSF 2012CB825601 at SDUWH, NNSFC Y2503BA110 at NSSC,
NSF grant AGS-0962653 and contract SA4889-26309 from the University of California Berkeley at APL,
and NSF grant AGS-0962666 at SWRI. We thank the referee for very useful suggestions.

\end{document}